\documentstyle[aps,prbbib,twocolumn]{revtex}

\begin{document}
\draft

\title{Frequency dependent admittance of a two-dimensional quantum wire}

\author{Jian Wang$^{(1)}$ and Hong Guo$^{(2)}$}

\address{(1) Department of Physics, The University of Hong Kong,
Pokfulam Road, Hong Kong.\\
(2) Centre for the Physics of Materials and
Department of Physics, McGill University,
Montr\'eal, Qu\'ebec, H3A 2T8 Canada.}

%\date{\today}
\maketitle

\begin{abstract}
The frequency dependent conductance of a two-dimensional quantum wire is
computed using a current conserving formalism.  The
correction to the dc-conductance due to a time-dependent potential 
is related to the local partial density of states which we compute
numerically. The current conservation is explicitly confirmed by
computing the global density of states and comparing it with a quantity
which is related to the electron dwell time.  Our calculation clearly
reveals the physical meaning of the various partial density of states.
\end{abstract}

\medskip


The physics associated with quantum conduction in various low
dimensional systems has been a major focus in current solid state
research. Due to advances in controlled crystal growth and lithographic
techniques, it is now possible to fabricate various two-dimensional
submicron structures where accurate experimental measurements can be
made\cite{review}.  On the theoretical side, the understanding of
quantum transport in these very small systems has been advanced by 
the Landauer theory\cite{landauer} of one-dimensional transport. In 
cases where multi-probes are attached to the conductor, B\"uttiker has
provided the useful formula for computing various
conductances\cite{buttiker1}.  The Landauer-B\"uttiker formalism has
been applied to a large class of quantum transport problems in the
mesoscopic and ballistic regimes where the external potential is static.
On the other hand, when external potential has a time-dependent
oscillating component, the Landauer-B\"uttiker formalism can not be
directly applied.  As shown by B\"uttiker and his
co-workers\cite{buttiker2} that a direct application of the original approach
of Landauer-B\"uttiker formalism, where the internal electric potential
distribution is not needed,  can not yield electric current and charge
conservation.  Hence the coherent quantum transport theory for mesoscopic
and ballistic conductors must be extended when time-varying external
potentials are to be included.  Clearly the problem of computing 
frequency dependent conductance for mesoscopic and ballistic conductors is
a very important subject from both theoretical and application points of
view.

Recently in a series of articles, B\"uttiker and 
co-workers\cite{buttiker2,but3,but4,but5} have developed a theoretical
formalism for investigating frequency-dependent coherent quantum transport 
for mesoscopic and ballistic conductors when electric potential in the
leads far away from the scattering region of the conductor is time varying.
This is an important advance in the theory of mesoscopic physics
and warrant further development of the theory and application.  In simple 
one dimensional systems such as a $\delta -$function potential, a perfect
quantum wire, or an 1D quantum well, the frequency dependent admittance
can be explicitly calculated
since the scattering matrix can be obtained analytically\cite{but5}.
However, in truly 2D systems the application of the formalism is less
straightforward. This is due partly to technical difficulties of computing
certain quantities such as the various partial density of states (PDOS, see
below).  These quantities, which appear naturally in the theoretical
formulation of the problem\cite{buttiker2}, have deep physical meaning which
is not obviously clear and needs to be revealed. To the best of our 
knowledge, there are so far no studies and direct calculations\cite{sarma} 
of the various partial density of states in 2D. But these quantities must be
obtained if the frequency dependent admittance is to be computed
from first principles. The purpose of this paper is to report 
our investigations on the dynamic transport properties of a 2D quantum 
wire system.

In particular, we consider a T-shaped quantum wire structure as shown in
Fig. (1).  The two probes extend to $x=\pm\infty$ while the scattering
region is provided by the T-shaped junction as shown being bounded by
the two dotted lines. This 2D quantum wire has been
studied by many authors\cite{hess,jian1} for the case of static transport.
While this is a simple 2D system, we found that much physical
intuition about various quantities can be obtained for the dynamic
transport (see below).  We have computed the frequency dependent
admittance for this system using the
current conserving formalism of Ref. \cite{buttiker2}. The
correction to the dc-conductance due to a time-dependent potential 
is related to the local partial density of states which we compute
numerically. The current conservation is explicitly confirmed by
computing the global density of states and comparing it with a quantity
which is related to the electron dwell time.  Our direct calculation clearly
reveals the physical meaning of the partial density of states which plays an
essential role in the dynamic transport theory.

For the sake of presentation we briefly review the current conserving 
dynamic transport formalism of B\"uttiker, Thomas, Pr\^etre, Gasparian 
and Christen\cite{buttiker2,but5}. Their theory proceeds in three steps.
The first step is to determine the current and charge density response of
the system to the time-variation of the external potential.  This leads 
to an admittance matrix with elements given by, to first order in $\omega$,
\begin{equation}
g_{\alpha\beta}^e(\omega )=g_{\alpha\beta}^e(0)-i\omega e^2
(dN_{\alpha\beta}/dE)\ \ \ \ \ ,
\label{ge}
\end{equation}
where the first term on the right hand side is the 
admittance (conductance) when frequency $\omega=0$ and this term is given by
the usual B\"uttiker multi-probe conductance formula\cite{buttiker1}. 
The second term gives a
correction due to the time-dependence of the external potential, and is
determined by the global partial density of states\cite{but5} (global PDOS),
$dN_{\alpha\beta}/dE$. Here the indices $\alpha\beta$ denote
scattering from a probe labeled by $\beta$ to that labeled by $\alpha$.
This admittance matrix $g_{\alpha\beta}^e(\omega )$ does not conserve 
current, since $\sum_{\alpha}g_{\alpha\beta}^e(\omega )\neq 0$.  
To correct this problem, one must consider the internal potential
distribution induced by the external perturbation.  This is computed in the
second step of the theory\cite{buttiker2}, and the result is rather simple
if only Thomas-Fermi linear screening is included. In the final step, one
computes the currents in the probes induced by this internal 
potential distribution, which leads to an additional term in the admittance
matrix.  Thus the total admittance is given by $g^I = g^e + g^i$, with
\begin{equation}
g_{\alpha\beta}^i\ =\ i\omega e^2D_{\alpha\beta}
\label{gi}
\end{equation}
and
\begin{equation}
D_{\alpha\beta}\ \equiv\ \int d{\bf r}
\left[\frac{dn(\alpha,{\bf r})}{dE}\right]
\left[\frac{dn({\bf r})}{dE}\right]^{-1}
\left[\frac{dn({\bf r},\beta)}{dE}\right]\ \ \ .
\label{dab}
\end{equation}
In Eq. (\ref{dab}) the quantity $dn({\bf r},\alpha)/dE$ is the 
{\it injectivity} which measures the additional local
charge density brought into the sample by the oscillating chemical potential
at probe $\alpha$. Without a magnetic field, the {\it emissivity} 
$dn(\beta,{\bf r})/dE$ equals to the injectivity. It was shown\cite{but5}
that these quantities could be computed from the electron dwell time or 
the scattering Green function.  Finally, the total local density of states
is $dn({\bf r})/dE=\sum_{\alpha}dn(\alpha,{\bf r})/dE$. 
With these expressions, we obtain the final formula for the frequency
dependent admittance to linear order in $\omega$,
\begin{equation}
g_{\alpha\beta}^I(\omega )\ =\ g_{\alpha\beta}^e(\omega=0 )
\ -\ i\omega e^2\left(\frac{dN_{\alpha\beta}}{dE}-D_{\alpha\beta}\right)
\ \ .
\label{gfinal}
\end{equation}
It is now straightforward 
to prove that current is conserved since the admittance matrix $g^I$ satisfies
$\sum_{\alpha}g_{\alpha\beta}^I(\omega)=0$. This can be seen by realizing
that $\sum_{\alpha}dN_{\alpha\beta}/dE\equiv d\bar{N}_{\beta}/dE$ is the
injectance which is identical to $\sum_{\alpha}D_{\alpha\beta}$.

Let's apply this formalism to the T-shaped 2D conductor. The incident 
electron comes from probe 1, it scatters at the T-junction, and then 
reflects back to probe 1 or transmits to probe 2. To be concrete we have
fixed the wire width to be $W$, the width and height of the side stub are
fixed at $aW$ and $bW$ with $a=b=1$. The units are fixed by $\hbar^2/2m=1$
with $m$ the effective mass of the electron, and lengths measured in terms
of $W$. For simplicity we have focused on the first {\it transport} subband 
only, thus the incoming electron energy is restricted: 
$(\pi/W)^2<E<(2\pi/W)^2$. In this case elements of the scattering 
matrix ${\bf s}_{\alpha\beta}$ are simple complex numbers. With more 
than one subbands they become matrices in the subband space. However 
the computation proceeds in similar fashion with one or more than one 
subbands. We have solved the quantum scattering problem using a mode 
matching method\cite{jian1}, where the wavefunction in the scattering region
was expanded using a suitable basis set and in particular $50$ modes were
included.  We have checked the convergence that including more modes in the
scattering region essentially does not change the results.  The scattering
probabilities were obtained by matching the wavefunctions and their
derivatives at the boundaries between the scattering region and the
probes.

To compute the admittance we need to know various partial density of states.
These quantities turn out to have very interesting and physically clear
property once they are plotted. The global PDOS is related to the
scattering matrix, and is approximately given by\cite{buttiker2,but5}
\begin{equation}
\frac{dN_{\alpha\beta}}{dE}\ =\ \frac{1}{4\pi i}
\left(s_{\alpha\beta}^{\dagger}\frac{ds_{\alpha\beta}}{dE}\ -\  
\frac{ds_{\alpha\beta}^{\dagger}}{dE}s_{\alpha\beta}\right)\ \ \ .
\label{gpdos}
\end{equation}
For a finite scattering volume there are corrections to this expression of 
the order $O(\frac{\lambda}{L})$ where $\lambda$ is the electron wavelength 
and $L$ the system size\cite{but5}. For large system sizes or large 
energies, Eq. (\ref{gpdos}) is adequate.
Later when we check the current conservation, the correction term will be
added. Since in 2D the expression for the scattering matrix can not be written
down analytically, we have decided to perform the energy derivatives of
(\ref{gpdos}) numerically. In particular we computed $s_{\alpha\beta}(E)$ at
various values of energy $E$ and used a $5$-point numerical derivative
to find $ds_{\alpha\beta}/dE$. The behavior of the global PDOS is plotted in
Fig. (2) together with the transmission coefficient $T_{21}(E)$. As our 
quantum wire system is very transmissive, $T_{21}$ has large values in 
general except at the two resonance levels where $T=0$.  
In a previous work we have shown that
the energies where $T=0$ corresponds to the quasi-bound states located in the
T-junction\cite{jian1}.  The behavior of the global partial density of states
coincides well with that of the transmission coefficients. In particular
$dN_{11}/dE$, which is the global PDOS for reflection, peaks at the
energies where transmission coefficient is minimum or reflection maximum.  On
the other hand, $dN_{21}/dE$, which is the global PDOS for transmission, 
takes minimum value when $T$ is minimum. Indeed, a general discussion for 
an one-dimensional system with a symmetric scatterer\cite{but5} reveals  
that $dN_{11}/dE\sim RdN/dE$ while $dN_{21}/dE\sim TdN/dE$ with $R$ and 
$T$ the reflection and transmission coefficients, and $dN/dE$ the global
DOS.  Hence these quantities have vivid physical meaning.  

The next quantity of interest is $D_{\alpha\beta}$ given by Eq. (\ref{dab}).
To compute this quantity, we first find the injectivity which is given by the
scattering wavefunction\cite{buttiker2} in the T-junction, at zero 
temperature this is given by
\begin{equation}
\frac{dn({\bf r},\alpha)}{dE}\ =\ \frac{1}{hJ}|\psi({\bf r})|^2
\label{injectivity}
\end{equation}
where $J$ is the incoming particle flux. Clearly a spatial integration of
this quantity gives the electron dwell time\cite{but6,jian1}
$\tau_{\alpha}$.
In general $D_{\alpha\beta}$ is obtained with two calculations of the 
wavefunction for particles coming from left and from right. Fig. (3) shows
this quantity as a function of the electron energy.  Both $D_{11}(E)$ and
$D_{21}(E)$ peak at the energies where the transmission takes the minimum
values. This is understandable since these special energies correspond to the
scattering states\cite{jian1} where the electron dwell time takes maximum
value, and the injectivity represents essentially the electron dwelling in 
an area $d{\bf r}$ at position ${\bf r}$ irrespective where it is finally
scattered.

With the global partial density states and the quantity $D_{\alpha\beta}$
calculated, we have thus obtained the admittance $g_{\alpha\beta}^I(\omega )$
from Eqs. (\ref{ge}), (\ref{gi}), and (\ref{gfinal}).
Fig. (4) shows the $\omega$-dependent part of $g^I$, 
$\hat{g}_{\alpha\beta}\equiv dN_{\alpha\beta}/dE -D_{\alpha\beta}$ as a 
function of energy $E$. Apart from the prefactor $\omega e^2$, this quantity 
$\hat{g}$ is the imaginary part of the admittance, and is called 
{\it emittance}. At the quasi-bound state
levels where transmission coefficient $T_{21}(E)$ takes minimum values,
$\hat{g}_{\alpha\beta}(E)$ also takes extremal values.  
It is interesting to observe that the transmissive part $\hat{g}_{21}$ takes
minimum value where $T(E)=0$, while the reflective part $\hat{g}_{11}$
takes maximum value at the same energy. Thus in this sense, the dynamic
part of the admittance $\hat{g}_{\alpha\beta}$ has the same behavior as the
static admittance $g_{\alpha\beta}(\omega=0)$ as a function of energy.
Furthermore, $\hat{g}_{\alpha\beta}$ changes sign as energy is varied: the
system responds either capacitively when 
$\hat{g}_{11}=-\hat{g}_{21}=-\hat{g}_{12}>0$, 
or inductively otherwise. Hence at the resonance (where $T\approx 0$ 
for our system) $\hat{g}_{21}$ and $\hat{g}_{11}$ are capacitive.
Fig. (4) shows the clear crossover between the capcitive and inductive
responses for this 2D quantum wire as energy is changed.  
%In contrast, for a double barrier system the response is inductive at 
%the resonance\cite{buttiker2}. However in that case the resonance is 
%marked by transmission coefficient being close to unity.

We can now explicitly confirm the current conservation 
by summing up the admittance matrix elements
and check whether or not $\sum_{\alpha}g_{\alpha\beta}^I = 0$.  Since the
global PDOS obtained using Eq. (\ref{gpdos}) is not exact, a correction
should be added. For an 1D system Gasparian {\it et. al.}\cite{gasparian}
have shown that
\begin{equation}
\frac{dN_{\beta}}{dE}\ \equiv\ \sum_{\alpha}\frac{dN_{\alpha\beta}}{dE}
\ +\ Im\left(\frac{s_{\beta\beta}}{4\pi E}\right)\ \ ,
\label{conservation}
\end{equation} 
where $E$ is the electron energy.  Unfortunately for a 2D system 
such as ours, the Green's function can not be written 
down analytically hence how to derive a similar correction 
term as that in (\ref{conservation}) is unclear. However
since we have numerical results of all the quantities, a reasonable 
correction term can easily be obtained. We found that the same 
form as (\ref{conservation}) led to almost perfect current 
conservation, provided we use the {\it transport}
energy $k^2$ as the energy $E$ in (\ref{conservation}).  Our data
in Fig. (5) clearly and unambiguously shows that the following is established
\begin{equation}
\sum_{\alpha}\frac{dN_{\alpha\beta}}{dE}
\ +\ Im\left(\frac{s_{\beta\beta}}{4\pi k^2}\right)\ =\
\sum_{\alpha}D_{\alpha\beta}\ \ .
\label{final}
\end{equation}
With this result, we have thus explicitly shown\cite{foot1} 
the current conservation $\sum_{\alpha}g_{\alpha\beta}^I=0$. Since the 
correction term is inversely proportional to $k^2$, it plays a role 
only at small $k$. This can already be seen in Fig. (4) where the two
curves add up to zero except at low energies.

In summary we have, for the first time, implemented 
the current conserving dynamic conductance formalism 
for 2D metallic conductors. In the simple case of
applying Thomas-Fermi linear screening for the 
interacting electrons (hence the analysis is more suitable for
metallic samples), the current response to the internal 
potential can be computed from the electron
dwell time using the scattering wavefunctions. The frequency dependent 
admittance of a T-shaped 2D quantum wire is calculated to linear order in
frequency.  We have explicitly confirmed the electric current conservation 
and found that a correction to the total density of states is needed at low
energies, in the same fashion as that of 1D case.  It is very interesting to
clarify the physical meaning of the partial DOS.  These quantities provide
information on the density of carriers {\em and} the transmission of
carriers from one contact to another.  At quantum resonances these
quantities take extremal values. There are many further applications
of this important theoretical formalism, to situations involving multi-mode,
magnetic fields, and non-linear screening. Another very challenging 
extension is to push the theory to higher order in frequency\cite{but5} and
formulate the theory in a way that permits a computational implementation.
We hope to be able to report these results in the future. 

\section*{Acknowledgments}

We thank Prof. M. B\"uttiker for helpful communications and discussions.
We gratefully acknowledge support by a RGC grant from the Government of 
Hong Kong under grant number HKU 261/95P, the Natural Sciences and 
Engineering Research Council of Canada and le Fonds pour la Formation de
Chercheurs et l'Aide \`a la Recherche de la Province du Qu\'ebec.
We thank the Computer Center of the University of Hong Kong for
computational facilities.

\begin{figure}
\caption{Schematic plot of the T-shaped quantum wire.  The wire width is 
$W$, the side-stub width and height is $aW$ and $bW$. The two dotted lines
separate the scattering region from the two probes.}
\end{figure}

\begin{figure}
\caption{
Global partial density of states and the transmission
coefficient as functions of electron energy $E$. Solid line: transmission 
coefficient $T_{21}$; dotted line: $dN_{11}/dE$;  dashed line: $dN_{21}/dE$.
Unit of energy is $\hbar^2/(2mW^2)$.}
\end{figure}

\begin{figure}
\caption{The current response to the internal potential,
$D_{\alpha\beta}$, as a function of energy $E$.  Solid line: $D_{11}$; 
dotted line: $D_{21}$.}
\end{figure}

\begin{figure}
\caption{The imaginary part (dynamic part) of the admittance,
$\hat{g}_{\alpha\beta}\equiv dN_{\alpha\beta}/dE-D_{\alpha\beta}$ as a
function of energy.}
\end{figure}

\begin{figure}
\caption{Explicit confirmation of the current conservation, 
Eq. (\ref{final}).
Solid line: $dN_{\beta}/dE$ given by Eq. (\ref{conservation}) and 
$\beta =1$; dotted line: $\tau_{\beta}/h\ =\ \sum_{\alpha}D_{\alpha \beta}$
with $\beta=1$.}
\end{figure}


\begin{references}
\bibitem{foot1}
Near the second subband threshold, there seems to be a small but systematic 
deviation to the equal sign of Eq. (\ref{final}). This deviation could be 
due the fact that transmission probability has sharp changes in its value 
near the onset of the second subband and Eq. (\ref{final}) does not include
multi-propagating modes.
\end{references}
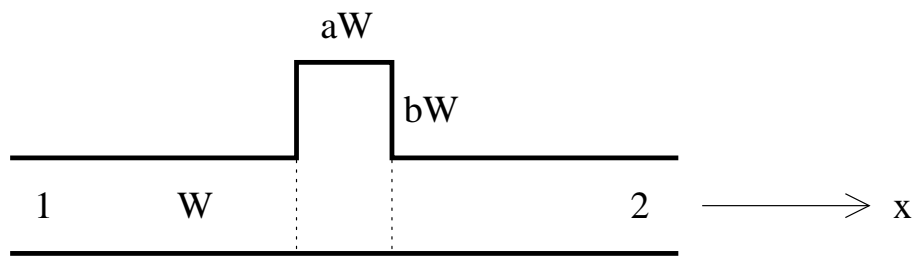
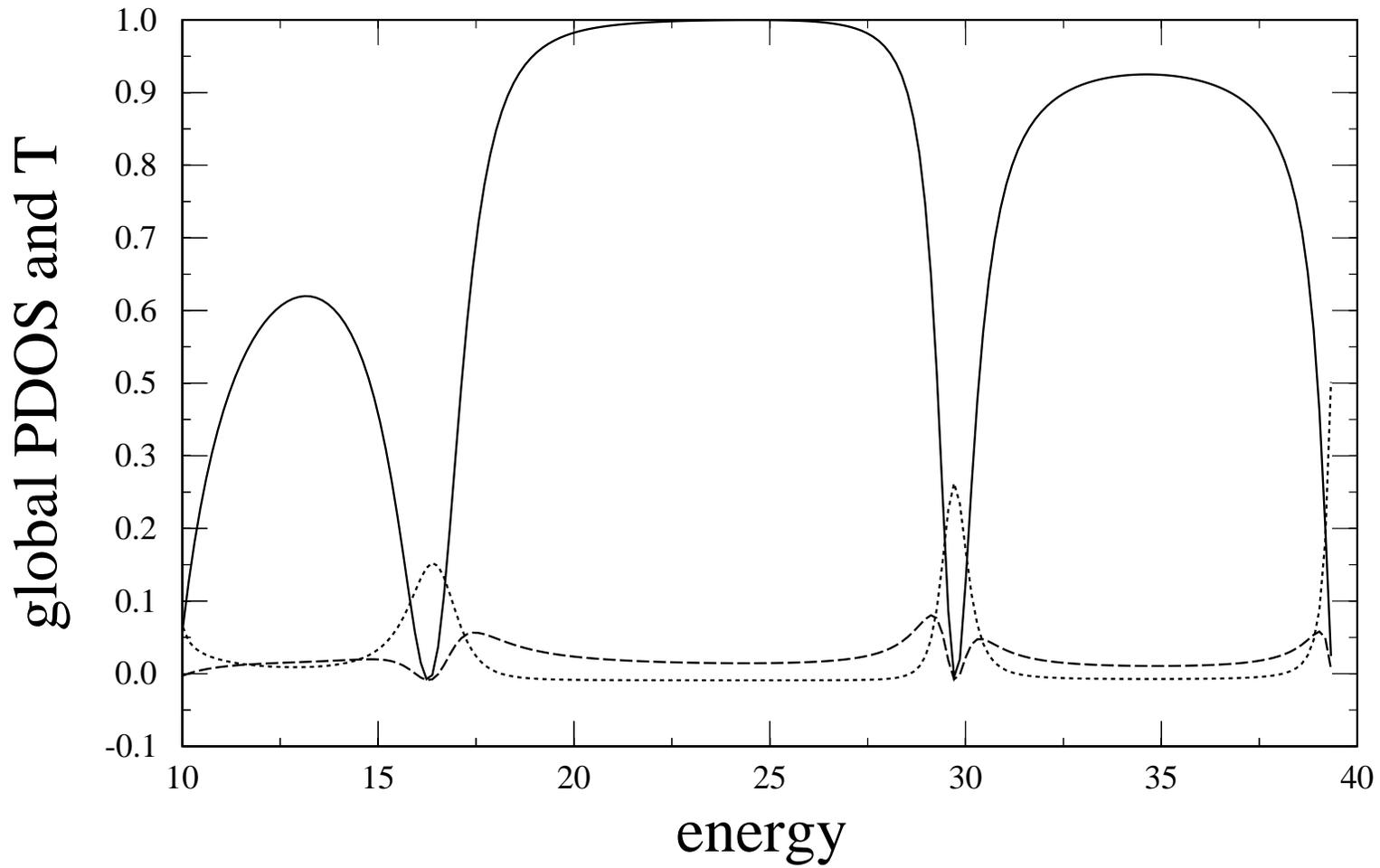
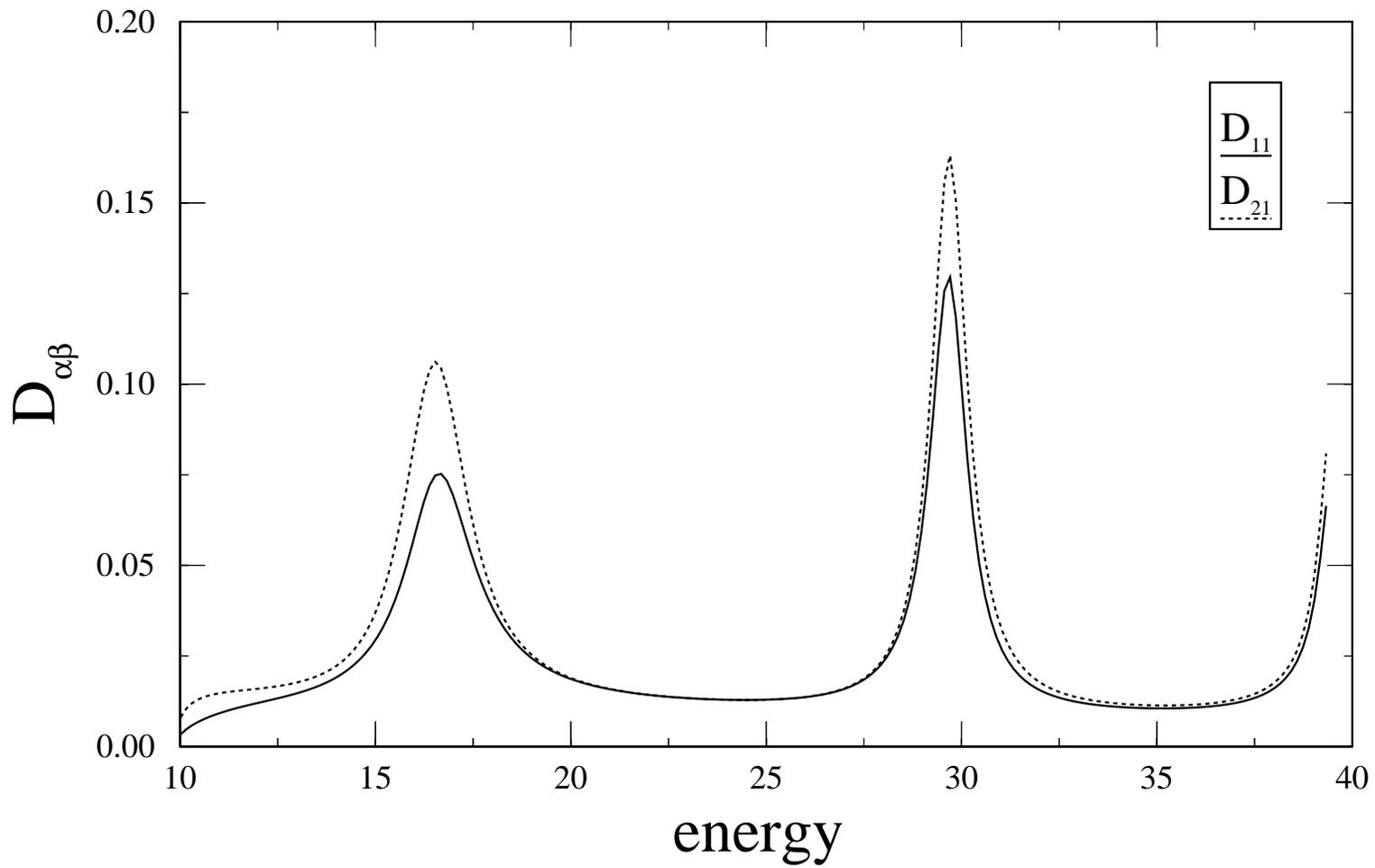
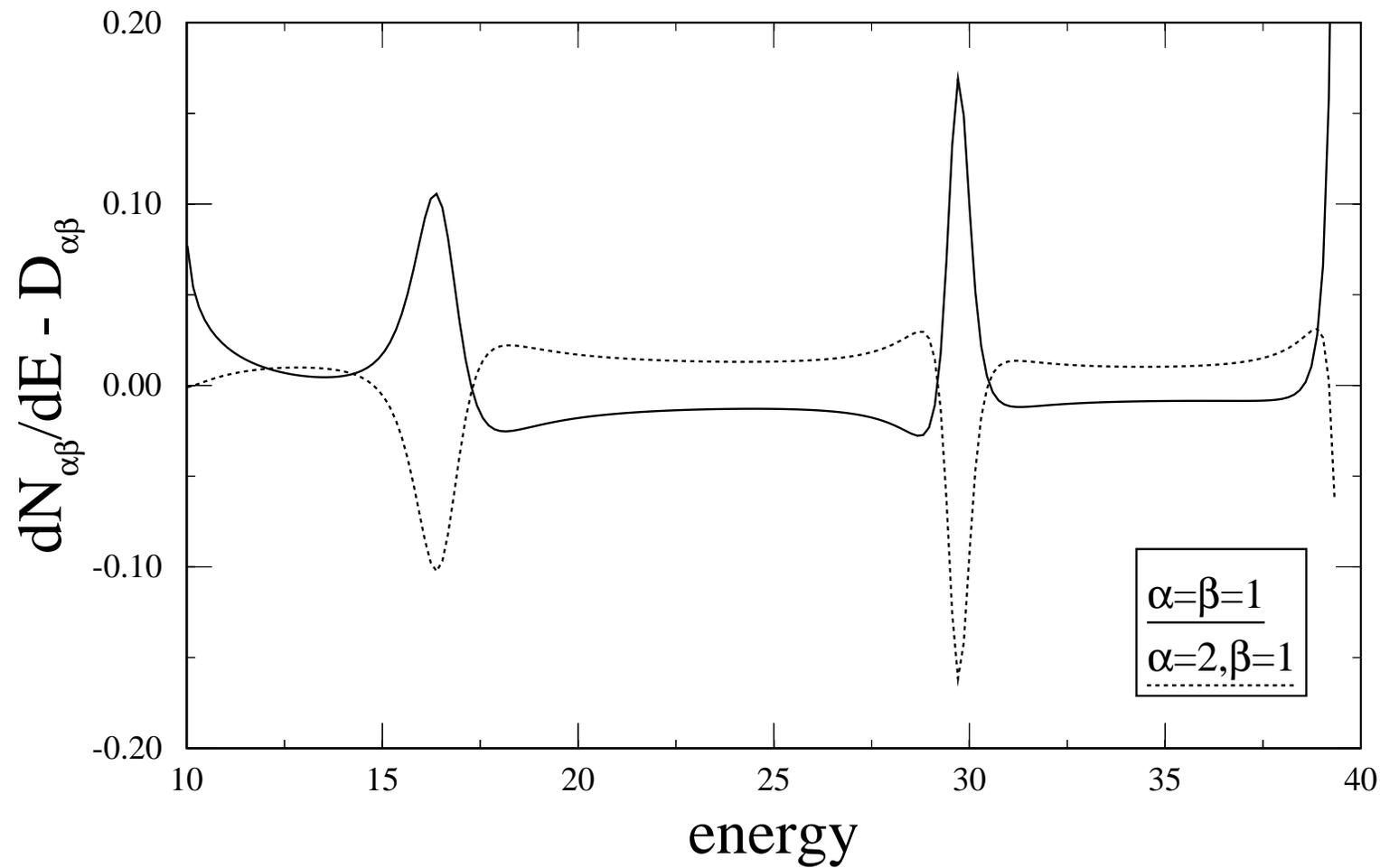
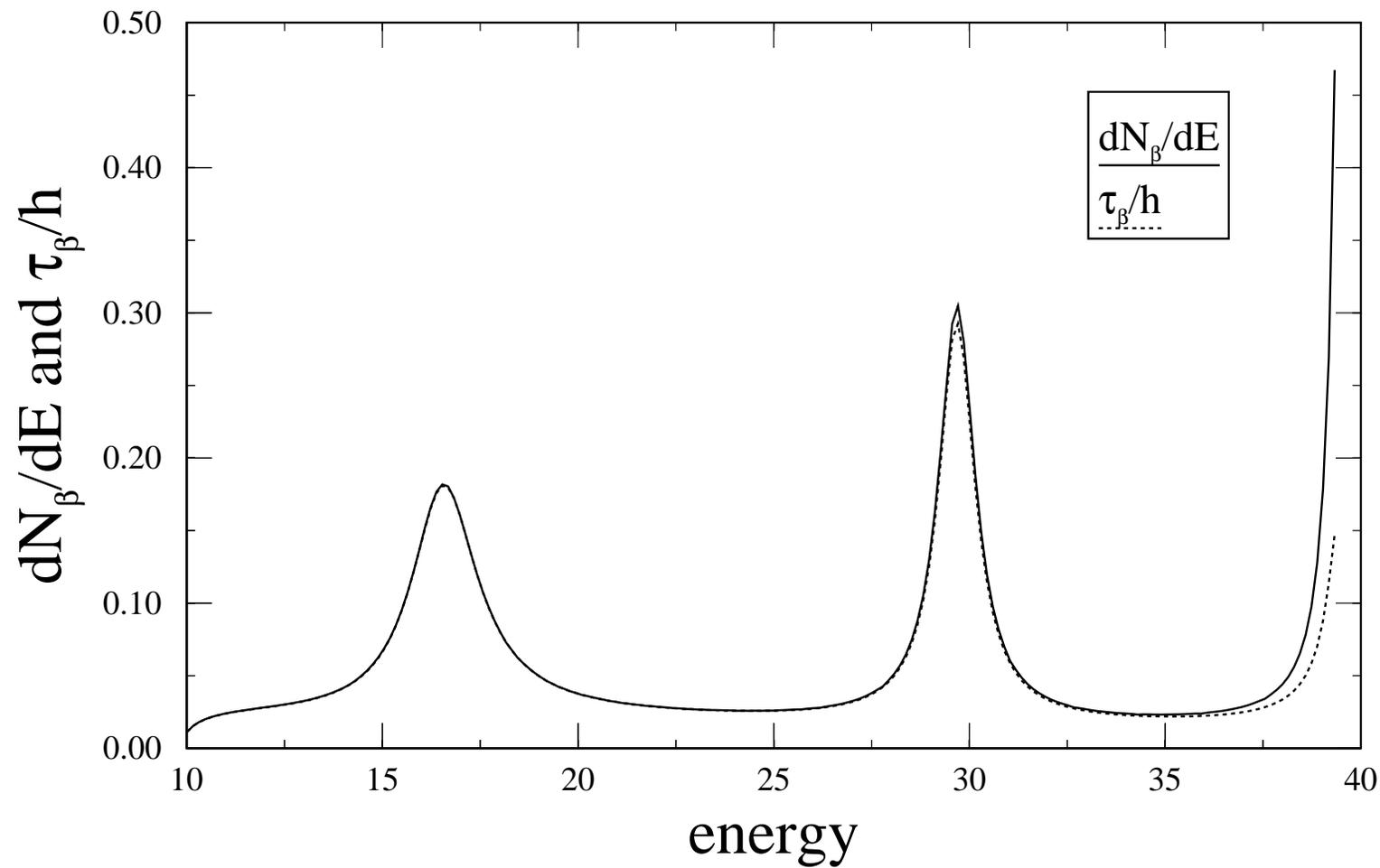
\end{document}